\documentstyle[12pt]{article}
\begin{document}
\pagestyle{empty} \setlength{\footskip}{2.0cm}
\setlength{\oddsidemargin}{0.5cm} \setlength{\evensidemargin}{0.5cm}
\renewcommand{\thepage}{-- \arabic{page} --}
\vspace*{-2.5cm}
\begin{flushright}
TOKUSHIMA 95-05 \\ (hep-ph/9511224) \\ November 1995
\end{flushright}
\vspace*{1.25cm}

\renewcommand{\thefootnote}{*)}
\centerline{\large\bf Structure of Electroweak Radiative
Corrections\,\footnote{Lecture presented at the XIX International
School on Theoretical Physics ``{\it Particle Physics and
Astrophysics in the Standard Model and Beyond}, Bystra, Poland,
September 19-26, 1995.}}

\vspace*{1.75cm}
\renewcommand{\thefootnote}{**)}
\centerline{\sc \phantom{**)}Zenr\=o HIOKI\,\footnote{E-mail:
hioki@ias.tokushima-u.ac.jp}}

\vspace*{1.75cm}
\centerline{\sl Institute of Theoretical Physics,\ University of
Tokushima}

\vskip 0.3cm
\centerline{\sl Tokushima 770,\ JAPAN}

\vspace*{3cm}
\centerline{ABSTRACT}

\vspace*{0.4cm}
\baselineskip=20pt plus 0.1pt minus 0.1pt
Looking into the inside of radiative corrections is an interesting
subject as a deeper study of the standard electroweak theory after
its remarkable success in the precision analyses. I will discuss here
a test of ``structure" of the EW radiative corrections to the
weak-boson masses, and show that we can now analyze several different
parts separately.
\vfill
\newpage
\renewcommand{\thefootnote}{\sharp\arabic{footnote}}
\renewcommand{\theequation}{\arabic{section}.\arabic{equation}}
\pagestyle{plain} \setcounter{footnote}{0}
\baselineskip=21.0pt plus 0.2pt minus 0.1pt
\section*{\normalsize\bf$\mbox{\boldmath $\S$}\!\!$
1. Introduction}
\setcounter{section}{1}\setcounter{equation}{0}

\vspace*{-0.3cm}
The standard electroweak theory has been excellently successful in
describing a lot of low- and high-energy precision data, by taking
into account radiative corrections (see \cite{EFL,PreAna} and
references cited therein). This means that the theory has been well
tested as a renormalizable field theory. Looking into the inside of
the EW radiative corrections is an interesting theme as one of its
next-step studies. At this School, I would like to discuss
``structure" of the EW corrections to the $W$ and $Z$ masses based on
my recent work \cite{ZH95}.

There is no room for an objection on using $M_Z^{exp}(=91.1884\pm
0.0022$ GeV \cite{LEP}), while the reason why I focus on the $W$ mass
among others is as follows: First of all, the weak-boson mass
relation derived from the radiative corrections to $G_F$ (the
$M_W$-$M_Z$ relation) has the advantage of being freest from gluon
effects. In addition, all the other high-energy precision data are
those on $Z$ boson or those at $\sqrt{s}\simeq M_Z$, and their
accuracy is now reaching the highest level, while $M_W$, which is
already known with a good precision, will be determined much more
precisely at LEP II. For comparison, the present $Z$ width is
${\mit\Gamma}_Z^{exp}=2.4963\pm 0.0032$ GeV \cite{LEP}, i.e.,
$\pm$0.13 \% precision. On the other hand, $M_W^{exp}$ by UA2+CDF+D0
is $80.26\pm 0.16$ GeV ($\pm$0.20 \% precision) \cite{wmass}, i.e.,
already comparable to ${\mit\Gamma}_Z^{exp}$, and its precision
reaches $\pm$0.06 \% once ${\mit\Delta}M_W^{exp}=\pm$50 MeV is
realized at LEP II \cite{Kaw}. Therefore, we can expect very clean
and precise tests through the $M_{W,Z}$ measurements.

I wish to proceed as follows: First of all, I will explain what I
mean by ``structure of $\cdots$" and what we should do in order to
test it in section 2. Then, a brief review of the EW corrections to
the weak-boson masses is given in $\S\,$3. In $\S\,$4, fermionic
corrections are studied. What I study there are the
(QED-)improved-Born approximation and the non-decoupling top-quark
effects. Testing the latter one is particularly important because the
existence of such effects is a characteristic feature of theories in
which particle masses are produced through spontaneous symmetry
breakdown plus large Yukawa couplings. In $\S\,$5, on the other hand,
I will study other-type corrections from $W$, $Z$ and the Higgs,
i.e., bosonic contributions. Since the top quark was found to be very
heavy \cite{TOP}, we have a good chance to detect the bosonic
contribution. This is because the fermionic leading-log terms and the
non-decoupling top-quark terms work to cancel each other, and
consequently the role of the non-fermionic corrections becomes
relatively more significant. The final section is for a summary and
brief discussions.
\section*{\normalsize\bf$\mbox{\boldmath $\S$}\!\!$
2. What ``Structure $\cdots$ " Means}
\setcounter{section}{2}\setcounter{equation}{0}

\vspace*{-0.3cm}
EW radiative corrections to physical quantities consist of several
parts with different properties. For example, one-loop corrections to
the muon-decay amplitude are usually expressed as ${\mit\Delta}r$,
and can be written as follows:
\begin{eqnarray}
{\mit\Delta}r={\mit\Delta}\alpha+{\mit\Delta}r[m_t]+
{\mit\Delta}r[m_{\phi}]+{\mit\Delta}r[\alpha]. \label{eq21}
\end{eqnarray}
Here ${\mit\Delta}\alpha$ is the leading-log terms from the light
charged fermions
\begin{eqnarray}
{\mit\Delta}\alpha=-\frac{2\alpha}{3\pi}\sum_{f(\neq t)}
\Bigl\{Q_f^2\ln\Bigl(\frac{m_f}{M_Z}\Bigr)+{5\over 6}\Bigr\},
\label{eq22}
\end{eqnarray}
${\mit\Delta}r[m_t]$ and ${\mit\Delta}r[m_{\phi}]$ express the
non-decoupling top-quark and Higgs-boson effects respectively
\begin{eqnarray}
{\mit\Delta}r[m_t]=-{\alpha\over{16\pi{\sl s}_W^2}}
\biggl\{ {3\over{{\sl s}_W^2 M_Z^2}}m_t^2
+4\biggl({{\sl c}_W^2\over{\sl s}_W^2}-{1\over 3}
-{{3m_b^2}\over{{\sl s}_W^2 M_Z^2}}\biggr)
\ln\Bigl({m_t\over M_Z}\Bigr)\biggr\},
\end{eqnarray}
\begin{eqnarray}
{\mit\Delta}r[m_{\phi}]={{11\alpha}\over{24\pi{\sl s}_W^2}}
\ln\Bigl({m_{\phi}\over M_Z}\Bigr),
\end{eqnarray}
where ${\sl c}_W^{\phantom 2}\equiv M_W/M_Z$ and ${\sl s}_W^2=1-
{\sl c}_W^2$, and ${\mit\Delta}r[\alpha]$ is the remaining
$O(\alpha)$ non-leading terms.

What I have so far studied is to see by using experimental data if
each of them must exist or not. I am afraid, however, this statement
will not be enough clear. From a purely theoretical point of view, it
may seem to be stupid to ask, e.g., if the data need the bosonic
effects. Everyone knows that $W^{\pm}$ and $Z$ exist, and since we
are studying in the framework of renormalizable field theories, their
loop effects must of course exist. Then, how about the Higgs
contribution? This may also sound a meaningless question. If it would
not exist, the theory becomes non-renormalizable, and the precision
analyses performed so far must face immediately a quite serious
difficulty. More generally, it is easy in many cases to judge
pure-theoretically if some terms under consideration are
necessary or not. That is, removing the corresponding terms would
break some symmetries and/or renormalizability. In this sense, we can
say that they must exist.

{}From a phenomenological point of view, however, it is totally a
different story. As an example, let us consider the meaning of
testing the triple gauge-boson couplings. Also in this case, it will
not be meaningful pure-theoretically, since if the size of the
coupling differs from the one predicted by the gauge principle, the
theory becomes again non-renormalizable. In other words, the success
of the electroweak theory in precision analyses means that all the
couplings are already known. Nevertheless, testing these couplings
is a very significant phenomenological analysis. We need to observe
them directly in order for the theory to be established. Testing the
neutral current structure has also a quite similar significance.
These show the reason why I believe studying the structure of the EW
corrections are indispensable.

Finally, let me summarize what we have to do in actual analyses.
Suppose we are trying to test in a theory the existence of some
effects phenomenologically. Then, we have to show that the following
two conditions are simultaneously satisfied:
\begin{itemize}
\item The theory cannot reproduce the data without the terms under
   consideration, no matter how we vary the remaining free
   parameters.
\item The theory can be consistent with the data by adjusting the
   free parameters {\it appropriately} (i.e., within experimentally-
   and theoretically-allowed range), once the corresponding terms are
   taken into account.
\end{itemize}
Needless to say, we have to have data and theoretical calculations
precise enough to distinguish these two clearly. In those analyses
it is safer to be conservative: That is, when we check the first
criterion, the less we rely on data, the more certain the result is.
On the contrary, for checking the second criterion, it is most
trustworthy if we can get a definite conclusion after taking into
account all the existing data, preliminary or not.
\section*{\normalsize\bf$\mbox{\boldmath $\S$}\!\!$
3. Corrections to the Weak-Boson Masses}
\setcounter{section}{3}\setcounter{equation}{0}

\vspace*{-0.3cm}
Through the $O(\alpha)$ corrections to the muon-decay amplitude,
the $W$ mass is calculated as
\begin{eqnarray}
M_W=M_W(\alpha, G_F, M_Z, {\mit\Delta}r). \label{eq31}
\end{eqnarray}
The explicit expression of Eq.(\ref{eq31}) at one-loop level with
resummation of the leading-log terms is
\begin{eqnarray}
M_W={1\over\sqrt{2}}M_Z
\biggl\{ 1+
\sqrt{\smash{1-{{2\sqrt{2}\pi\alpha}
\over{M_Z^2 G_F (1-{\mit\Delta}r)}}}
\vphantom{A^2\over A}
}~\biggr\}^{1/2}. \label{eq32}
\end{eqnarray}

When we apply the first criterion mentioned in the previous section
to the fermionic corrections, this formula is enough precise.
However, over the past several years, some corrections beyond the
one-loop approximation have been computed to it. They are two-loop
top-quark corrections and QCD corrections up to
$O(\alpha_{\rm QCD}^2)$ for ${\mit\Delta}r[m_t]$, and
$O(\alpha_{\rm QCD})$ corrections for the non-leading terms
\cite{BBCCV,HKl} (see \cite{FKS} as reviews). As a result, we have
now a formula including $O(\alpha\alpha_{\rm QCD}^2 m_t^2)$ and
$O(\alpha^2 m_t^4)$ effects:
\begin{eqnarray}
M_W=\sqrt{\rho\over 2}M_Z
\biggl\{ 1+
\sqrt{\smash{1-{{2\sqrt{2}\pi\alpha(M_Z)}
\over{\rho M_Z^2 G_F (1-{\mit\Delta}r_{rem})}}}
\vphantom{A^2\over A}
}~\biggr\}^{1/2}, \label{eq33}
\end{eqnarray}
\begin{eqnarray*}
&& \rho=1/(1-3\sqrt{2}G_F m_t^2/16\pi^2+{\mit\Delta}),\\
&& {\mit\Delta}r_{rem}=({\mit\Delta}r-{\mit\Delta}\alpha
    +3\sqrt{2}G_F c_W^2 m_t^2/16\pi^2 s_W^2 +{\mit\Delta}'),
\end{eqnarray*}
where ${\mit\Delta}$ and ${\mit\Delta}'$ are the above mentioned
higher-loop terms.

If ${\mit\Delta}r_{rem}$, the non-leading corrections, were to be
zero, Eq.(\ref{eq33}) would be unambiguous within the present
approximation. However, it is indeed not negligible. Concerning
how to handle it, there are several possible ways. I compute $M_W$
these several ways and use the average of the results as the central
value, while the difference among them is taken into account as part
of the theoretical error. This problem is discussed in detail in
\cite{BHP}. Anyway I use Eq.(\ref{eq31}) in the following to express
both Eqs.(\ref{eq32}) and (\ref{eq33}) for simplicity.

Let us see here what we can say about the whole radiative
corrections as a simple example of applications of the $M_W$-$M_Z$
relation and the two criterions given in $\S\,$2. Through
Eq.(\ref{eq31}), we have
\begin{eqnarray}
M_W^{(0)}=80.9400\pm 0.0027\ {\rm GeV\ \ and}\ \
M_W=80.36\pm 0.09\ {\rm GeV}
\end{eqnarray}
where $M_W^{(0)}\equiv M_W(\alpha, G_F, M_Z, {\mit\Delta}r=0)$ and
$M_W$ is for $m_t^{exp}=180 \pm 12$ GeV \cite{TOP}, $m_{\phi}=300$
GeV and $\alpha_{\rm QCD}(M_Z)$=0.118. Concerning the uncertainty of
$M_W$, 0.09 GeV, I have a little overestimated for safety.

As is easily found from Eq.(\ref{eq32}), $M_W^{(0)}$ depends only on
$\alpha$, $G_F$ and $M_Z$. So, we conclude from $M_W^{(0)}-M_W^{exp}=
0.68\pm 0.16$ GeV and $M_W-M_W^{exp}=0.10\pm 0.18$ GeV that
\begin{itemize}
\item $M_W^{(0)}$ is in disagreement with $M_W^{exp}$ at about
4.3$\sigma$ (99.998 \% C.L.),
\item $M_W$ is consistent with the data for, e.g., $m_{\phi}=300$
GeV, which is allowed by the present data $m_{\phi}>65.1$ GeV
\cite{Higgs}.
\end{itemize}
That is, the two criterions are both clearly satisfied, by which
the existence of radiative corrections is confirmed. Radiative
corrections were established at $3\sigma$ level already in the
analyses in \cite{Ama}, but where one had to fully use all the
available low- and high-energy data. We can now achieve a much higher
accuracy via the weak-boson masses alone. Analyses in the following
sections are performed in the same way as this, so I do not repeat
the explanation on the second criterion below since it is common to
all analyses.
\section*{\normalsize\bf$\mbox{\boldmath $\S$}\!\!$
4. Fermionic Corrections}
\setcounter{section}{4}\setcounter{equation}{0}

\vspace*{-0.3cm}
It is known that all the precision data up to 1993 are reproduced at
$1\sigma$ level by using $\alpha(M_Z)(=\alpha/(1-{\mit\Delta}\alpha))
$ instead of $\alpha$ in tree quantities \cite{NOV93}, where
$\alpha(M_Z)$ is known to be $1/(128.92\pm 0.12)$.\footnote{Recently
   three papers appeared in which $\alpha(M_Z)$ is re-evaluated from
   the data of the total cross section of $e^+ e^-\rightarrow\gamma^*
   \rightarrow hadrons$ \cite{alphamz} (the latest results are
   given in \cite{tatsu}). Here I simply took the average of the
   maximum and minimum among them.}\ 
I examine first whether this (QED-)Improved-Born approximation still
works or not.

The $W$ mass is calculated within this approximation as
\begin{eqnarray}
M_W[{\rm Born}](\equiv M_W(\alpha(M_Z),G_F,M_Z,0))
=79.963\pm 0.017\ {\rm GeV},
\end{eqnarray}
which leads to
\begin{eqnarray}
M_W^{exp}-M_W[{\rm Born}]~=~0.30\pm 0.16~{\rm GeV}.
\end{eqnarray}
This means that $M_W[{\rm Born}]$ is in disagreement with the data
now at $1.9\sigma$, which corresponds to about 94.3 \%\ C.L..
Although the precision is not yet sufficiently high,\footnote{The
   effective mixing angle in the $\bar{\ell}\ell Z$ vertex, $\sin^2
   \theta_{\ell}^{eff}$, is also a (almost) gluon-free quantity.
   Within this approximation, $\sin^2\theta_{\ell}^{eff}[{\rm Born}]$
   is given by $1-(M_W[{\rm Born}]/M_Z)^2=0.23105\pm 0.00033$. So,
   when $\sin^2\theta_{\ell}^{eff}=0.23186\pm 0.00034$ by LEP
   \cite{LEP} is taken into account, we will have a higher precision.
   In fact, the total $\chi^2$ becomes 6.58, which means that
   non-Born effects are required at 96.3 \% C.L.. However, when the
   SLD data via the LR-asymmetry are incorporated, the average
   becomes $\sin^2\theta_{\ell}^{eff}=0.23143\pm 0.00028$, and we can
   no longer get a better precision. This is why I did not use this
   quantity in my analysis.}\ 
it indicates some non-Born terms are needed which give a positive
contribution to the $W$ mass. It is noteworthy since the electroweak
theory predicts such positive non-Born type corrections unless the
Higgs is extremely heavy (beyond TeV scale). Similar analyses were
made also in \cite{NOV94}.

Next, I study the non-decoupling top-quark contribution. According to
my strategy, I computed the $W$ mass by using the following
${\mit\Delta}r'$ instead of ${\mit\Delta}r$ in Eq.(\ref{eq31}):
\begin{eqnarray}
{\mit\Delta}r'\equiv {\mit\Delta}r-{\mit\Delta}r[m_t].
\end{eqnarray}
The resultant $W$ mass is denoted as $M_W'$. The important point is
to subtract not only $m_t^2$ term but also $\ln(m_t/M_Z)$ term,
though the latter produces only very small effects unless $m_t$ is
extremely large. ${\mit\Delta}r'$ still includes $m_t$ dependent
terms, but no longer diverges for $m_t\to +\infty$ thanks to this
subtraction. I found that $M_W'$ takes the maximum value for the
largest $m_t$ and the smallest $m_{\phi}$ (as long as the
perturbation theory is applicable\footnote{We do not know what will
    happen for, e.g., $m_{\phi}=10$ TeV.}).\ 
That is, we get an inequality
\begin{eqnarray}
M_W'\ \leq\ M_W'[m_t^{max}, m_{\phi}^{min}].
\end{eqnarray}

We can use $m_t^{exp}=180\pm 12$ GeV \cite{TOP} and $m_{\phi}^{exp}
>65.1$ GeV \cite{Higgs} in the right-hand side of the above
inequality, i.e., $m_t^{max}=180+12$ GeV and $m_{\phi}^{min}=65.1$
GeV, but I first take $m_t^{max}\to +\infty$ and $m_{\phi}^{min}=0$
in order to make the result as data-independent as possible. The
accompanying uncertainty for $M_W'$ is estimated at most to be about
0.03 GeV. We have then
\begin{eqnarray}
M_W' < 79.950 (\pm 0.030) \ {\rm GeV\ \ and\ \ }
M_W^{exp}-M_W' > 0.31\pm 0.16\ {\rm GeV},
\end{eqnarray}
which show that $M_W'$ is in disagreement with $M_W^{exp}$ at about
$1.9\sigma$. This means that 1) the electroweak theory is not able to
be consistent with $M_W^{exp}$ {\it whatever values $m_t$ and
$m_{\phi}$ take} if ${\mit\Delta}r[m_t]$ does not exist, and 2) the
theory with ${\mit\Delta}r[m_t]$ works well, as shown before, for
experimentally-allowed $m_t$ and $m_{\phi}$.

Combining them, we can summarize that the latest experimental data of
$M_{W,Z}$ demand the existence of the non-decoupling top-quark
corrections. This shows that we could know something about the
existence of the top even if we would know nothing about $m_t$ and
$m_{\phi}$. Of course, it never means that the present information on
them is not useful: The confidence level of this result becomes
higher if we use $m_t^{max}=180+12$ GeV and $m_{\phi}^{min}=65.1$
GeV:
\begin{eqnarray}
M_W' < 79.863 (\pm 0.030) \ {\rm GeV\ \ and\ \ }
M_W^{exp}-M_W' > 0.40\pm 0.16\ {\rm GeV},
\end{eqnarray}
that is, $2.5\sigma$ level.
\section*{\normalsize\bf$\mbox{\boldmath $\S$}\!\!$
5. Corrections Including Bosonic Effects}
\setcounter{section}{5}\setcounter{equation}{0}

\vspace*{-0.3cm}
I wish to examine in this section non-fermionic contributions to
${\mit\Delta}r$ (i.e., the Higgs and gauge-boson contributions). It
has been pointed out in \cite{DSKK} by using various high-energy
data that such bosonic electroweak corrections are now inevitable. I
study here whether we can observe a similar evidence in the
$M_W$-$M_Z$ relation.

For this purpose, we have to compute $M_W$ taking account of only the
pure-fermionic corrections ${\mit\Delta}r[f]$. Since ${\mit\Delta}
r[f]$ depends on $m_t$ strongly, it is not easy to develop a
quantitative analysis of it without knowing $m_t$. Therefore, I used
$m_t^{exp}$ from the beginning in this case. I express thus-computed
$W$-mass as $M_W[{\rm f}]$. The result became
\begin{eqnarray}
M_W[{\rm f}]=80.48\pm 0.09\ {\rm GeV}.
\end{eqnarray}
This value is of course independent of the Higgs mass, and leads to
\begin{eqnarray}
M_W[{\rm f}]-M_W^{exp}=0.22\pm 0.18\ {\rm GeV},
\end{eqnarray}
which tells us that some non-fermionic contribution is necessary at
$1.2\sigma$ level. It is of course too early to say from this result
that the bosonic effects were confirmed. Nevertheless, this is an
interesting result since we could observe nothing before: Actually,
the best information on $m_t$ before the first CDF report (1994) was
the bound $m_t^{exp}>$ 131 GeV by D0 \cite{D0}, but we can thereby
get only $M_W[{\rm f}]>$ 80.19 ($\pm$0.03) GeV while $M_W^{exp}[94]$
was $80.23\pm 0.18$ GeV (i.e., $M_W[{\rm f}]-M_W^{exp} > -0.04\pm
0.18$ GeV).

For comparison, let us make the same computation for
${\mit\Delta}M_W^{exp}=\pm 0.05$ GeV and
${\mit\Delta}m_t^{exp}=\pm 5$ GeV, which will be eventually realized
in the future at Tevatron and LEP II. Concretely,
${\mit\Delta}m_t^{exp}=\pm 5$ GeV produces an error of $\pm 0.03$ GeV
in the $W$-mass calculation. Combining this with the theoretical
ambiguity ${\mit\Delta}M_W=\pm 0.03$ GeV, we can compute
$M_W[\cdots]-M_W^{exp}$ with an error of about $\pm 0.07$ GeV. Then,
$M_W[{\rm f}]-M_W^{exp}$ becomes $0.22\pm 0.07$ GeV if the central
value of $M_W^{exp}$ is the same, by which we can confirm the above
statement at $3\sigma$ level.

It must be very interesting if we can find moreover the existence of
the non-decoupling Higgs effects since we still have no
phenomenological indication for the Higgs boson. Then, can we in fact
perform such a test? It depends on how heavy the Higgs is: If it is
much heavier than the weak bosons, then we may be able to test it. If
not, however, that test will lose its meaning essentially, since
${\mit\Delta}r[m_{\phi}]$ comes from the expansion of terms like
$\int^1_0 dx\ln\{m^2_{\phi}(1-x)+M^2_Z x-M^2_Z x(1-x)\}$ in powers of
$M_Z/m_{\phi}$. Here, let us simply assume as an example that we have
gained in some way (e.g., at LHC) a bound $m_{\phi} > 500$ GeV. At
the same time, I assume ${\mit\Delta}M_W^{exp}=\pm 0.05$ GeV and
${\mit\Delta}m_t^{exp}=\pm 5$ GeV, since $M_W$ and $m_t$ will have
been measured at least at this precision by the time we get a bound
like $m_{\phi} > 500$ GeV. Then, for ${\mit\Delta}r''\equiv
{\mit\Delta}r-{\mit\Delta}r[m_{\phi}]$, the $W$ mass (written as
$M_W''$) satisfies $M_W'' > 80.46\pm 0.04$ GeV, where the
non-decoupling $m_{\phi}$ terms in the two-loop top-quark corrections
were also eliminated. This inequality leads us to $M_W''-M^{exp}_W >
0.20\pm 0.07$ GeV.

It seems therefore that we may have a chance to get an indirect
evidence of the Higgs boson even if future accelerators fail to
discover it.
\section*{\normalsize\bf$\mbox{\boldmath $\S$}\!\!$
6. Summary and Discussions}
\setcounter{section}{6}\setcounter{equation}{0}

\vspace*{-0.3cm}
A lot of experimental and theoretical effort has so far been made to
analyze the electroweak theory, and now we know that including the
radiative corrections is indispensable in these analyses. Based on
this success, I have carried out a further study of the theory and
its radiative corrections \cite{ZH95}, and reported here its main
results: They are analyses on (1) pure-fermionic and (2) bosonic
corrections in the weak-boson mass relation.

On the former part, I tested the improved-Born approximation and the
non-decoupling top corrections. There we could conclude that
non-Born type corrections and non-decoupling $m_t$ contribution are
required respectively at about $1.9\sigma$ and $2.5\sigma$ level by
the recent data on $M_{W,Z}$. This is a clean, though not yet
perfect, test of those corrections which has the least dependence on
hadronic contributions.

Concerning the latter part, we could observe a small indication for
non-fermionic contributions (at $1.2\sigma$ level), which can be
interpreted as the bosonic ($W/Z$ and the Higgs) corrections.
Furthermore, it seemed to be possible to test the non-decoupling
Higgs effects if the Higgs boson is heavy (e.g.,
$\lower0.5ex\hbox{$\buildrel >\over\sim$}$ 500 GeV). These results
(except for the last one) are visually represented in the Figure.

\centerline{\bf ------------------------}

\centerline{\bf Figure}

\centerline{\bf ------------------------}

On the bosonic corrections, however, supplementary discussions are
necessary. That is, the corresponding result is still somewhat
``unstable". I used in section 5 the present world average
$M_W^{exp}=80.26\pm 0.16$ GeV, but if the preliminary D0 data
$M_W^{exp}[{\rm D0}]=79.86\pm 0.40$ GeV and the early-stage CDF data
$M_W^{exp}[{\rm CDF(90)}]=79.91\pm 0.39$ GeV are not incorporated,
the average becomes $M_W^{exp}=80.40\pm 0.16$ GeV (CDF[92/93]+UA2).
This value might be more reliable, and in this case
\begin{eqnarray}
M_W[{\rm f}]-M_W^{exp}=0.08\pm 0.18\ {\rm GeV},
\end{eqnarray}
by which the bosonic effects become again totally unclear. On the
contrary, our conclusion on the fermionic corrections becomes thereby
much stronger: the non-Born effects and the non-decoupling $m_t$
effects are required respectively at 2.8$\sigma$ (99.5 \% C.L.) and
3.4$\sigma$ (99.9 \% C.L.).

More precise measurements of the top-quark and $W$-boson masses are
therefore considerably significant for studying this issue, and I
wish to expect that the Tevatron and LEP II will give us a good
answer for it in the very near future.

\vspace*{0.6cm}
\centerline{ACKNOWLEDGEMENTS}

\vspace*{0.3cm}
I am grateful to Marek Zra{\l}ek, Jan S{\l}adkowski and the
organizing committee of the School for their invitation and warm
hospitality. I also would like to thank Bohdan
Grz${{{\rm a}_{}}_{}}_{\hskip -0.18cm\varsigma}$dkowski for kindly
inviting me to Warsaw before this School, and Wolfgang Hollik for
correspondence on their work \cite{BHP}. During the School, I enjoyed
valuable conversations with many participants, which I appreciate
very much.

This work is supported in part by the Grant-in-Aid for Scientific
Research (No. 06640401) from the Ministry of Education, Science,
Sports and Culture, Japan.

\vspace*{0.3cm}

\newpage
\setlength{\unitlength}{0.1mm}          
\vspace*{-0.5cm}
\begin{picture}(1380,1500)(0,-100)
\put(-20,100){\line(0,1){1190}}
\put(-20,100){\vector(1,0){1370}}
\put(530,100){\line(0,1){1190}}
\multiput(850,100)(0,50){24}{\line(0,1){40}}
\put(600,100){\line(0,1){20}}
\put(1100,100){\line(0,1){20}}
\put(540,50){\makebox(100,50)[bl]{$-0.5$}}
\put(840,50){\makebox(100,50)[bl]{$0$}}
\put(1075,50){\makebox(100,50)[bl]{$0.5$}}
\put(780,1320){\makebox(500,50)[bl]{``happy" line}}
\put(700,-20){\makebox(500,50)[bl]{$M_W[\cdots]-M_W^{exp}$ (GeV)}}
\put(10,1200){\makebox(400,70)[bl]{$M_W^{(0)}$}}
\put(10,1140){\makebox(400,70)[bl]{($=80.9400\pm 0.0027$ GeV)}}
\put(1190,1200){\circle*{20}}
\put(1110,1200){\line(1,0){160}}
\put(10,1000){\makebox(400,70)[bl]{$M_W[{\rm Born}]$}}
\put(10,940){\makebox(400,70)[bl]{($=79.963\pm 0.017$ GeV)}}
\put(700,1000){\circle*{20}}
\put(780,1000){\line(-1,0){160}}
\put(10,800){\makebox(400,70)[bl]{$M_W'$}}
\put(10,740){\makebox(400,70)[bl]{($<79.950\pm 0.030$ GeV [a])}}
\put(695,750){\circle*{20}}
\put(775,750){\vector(-1,0){180}}
\put(10,680){\makebox(400,70)[bl]{($<79.863\pm 0.030$ GeV [b])}}
\put(650,690){\circle*{20}}
\put(730,690){\vector(-1,0){180}}
\put(10,540){\makebox(400,70)[bl]{$M_W[{\rm f}]$}}
\put(10,480){\makebox(400,70)[bl]{($=80.48\pm 0.09$ GeV)}}
\put(960,540){\circle*{20}}
\put(870,540){\line(1,0){180}}
\put(10,260){\makebox(400,70)[bl]{$M_W$}}
\put(10,200){\makebox(400,70)[bl]{($=80.36\pm 0.09$ GeV)}}
\put(900,260){\circle*{20}}
\put(810,260){\line(1,0){180}}
\end{picture}

\vspace*{0.5cm}
\centerline{\bf Figure}

Deviations of $W$ masses calculated in various approximations from
$M_W^{exp}=80.26\pm 0.16$ GeV. $M_W'[{\rm a}]$ is for
$m_t^{max}=+\infty$ and $m_{\phi}^{min}=0$ GeV, and $M_W'[{\rm b}]$
is for $m_t^{max}=192$ GeV and $m_{\phi}^{min}=65.1$ GeV. The last
$M_W$ (the one with the full corrections) is for $m_{\phi}=300$ GeV.
Only $M_W -M_W^{exp}$ crosses the ``happy" line.
\end{document}